\documentclass[twocolumn]{aastex63}

\usepackage[utf8]{inputenc}
\usepackage[toc,page]{appendix}
\usepackage{helvet}

\usepackage{fancyhdr}
\usepackage{amssymb}
\usepackage{amsmath}
\usepackage{natbib}
\usepackage{wrapfig}
\usepackage{graphicx}
\usepackage{mathrsfs}
\usepackage{booktabs}
\usepackage{lipsum}
\usepackage[font=Large,caption=false]{subfig}
\usepackage{mwe}
\usepackage{color,soul}

\shorttitle{Partial TDE Fallback Rates}
\shortauthors{P. R. Miles et al.}

\graphicspath{{./}{figures/}}

\begin{document}

\title{Fallback Rates from Partial Tidal Disruption Events}

\email{pmiles@syr.edu}

\author[0000-0003-1354-1984]{Patrick R. Miles}
\affiliation{Department of Physics, Syracuse University, Syracuse, NY 13244}

\author[0000-0003-3765-6401]{Eric R.~Coughlin}
\affiliation{Department of Astrophysical Sciences, Peyton Hall, Princeton University, Princeton, NJ 08544}
\affiliation{Department of Physics, Syracuse University, Syracuse, NY 13244}

\author[0000-0002-2137-4146]{C.~J.~Nixon}
\affiliation{Department of Physics and Astronomy, University of Leicester, Leicester, LE1 7RH, UK}

\begin{abstract}

A tidal disruption event (TDE) occurs when a star plunges through a supermassive black hole's tidal radius, at which point the star’s self-gravity is overwhelmed by the tidal gravity of the black hole. In a partial TDE, where the star does not reach the full disruption radius, only a fraction of the star’s mass is tidally stripped while the rest remains intact in the form of a surviving core. Analytical arguments have recently suggested that the temporal scaling of the fallback rate of debris to the black hole asymptotes to $t^{-9/4}$ for partial disruptions, effectively independently of the mass of the intact core. We present hydrodynamical simulations that verify the existence of this predicted, $t^{-9/4}$ scaling. We also define a break timescale -- the time at which the fallback rate transitions from a $t^{-5/3}$ scaling to the characteristic $t^{-9/4}$ scaling -- and measure this break timescale as a function of the impact parameter and the surviving core mass. These results deepen our understanding of the properties and breadth of possible fallback curves expected from TDEs and will therefore facilitate more accurate interpretation of data from wide-field surveys.

\end{abstract}


\section{Introduction} \label{sec:intro}
Supermassive black holes (SMBHs) are believed to reside at the centers of nearly every galaxy \citep{Kormendy_Richstone_1995}. Some actively consume surrounding gas through viscous accretion, whereby angular momentum in an accretion disk is transported outward in exchange for material being transported inward \citep{Lynden-Bell_Pringle_1974}. The viscous dissipation of energy is thought to be responsible for the extreme luminosities of these active galactic nuclei (AGN) \citep{salpeter64, Lynden-Bell_1969}.

Most SMBHs, however, emit little to no light \citep{Ho_2008} and their darkness can only occasionally be punctuated by a sudden flare that brightens and fades over months to years. These flashes are often attributed to tidal disruption events (TDEs) \citep{Rees_1988}, in which a star is ripped apart by the tidal field of a black hole. In a full disruption the star is completely destroyed, roughly half of the stellar debris stream remains gravitationally bound to the black hole, forms an accretion disk, and generates a bright flare. Partial disruptions may also occur in which only a fraction of the star's mass is tidally stripped by the SMBH, leaving behind a stellar core that survives the encounter. Both full (and, to a lesser extent, partial) TDEs have been the subject of extensive numerical study \citep[e.g.,][]{lacy82, Bicknell_Gingold_1983, Evans_Kochanek_1989, laguna93, LKP_2009, Hayasaki_2013, Guillochon_2013, tejeda13, Guillochon_2014, coughlin15, gafton15, Hayasaki_2016, Shiokawa_2015, Bonnerot_2016, Sadowski_2016, coughlin17, Wu_2018, Golightly_2019, Golightly_2019_spin, gafton19}, and observations of their resulting flares have been used to characterize otherwise-quiescent galactic nuclei \citep[e.g.,][]{komossa2015tidal, Komossa_2017, velzen2018science, Alexander_2017, Hung_2017, Holoien_2019, van_Velzen_2019,holoien20}.

Many characteristics of the light curves from TDEs are determined by the accretion rate of debris onto the black hole \citep[e.g.,][]{Lodato_Rossi_2011, Roth_2016}. This accretion rate is closely approximated by the rate at which debris returns to pericenter, known as the fallback rate, if (1) the kinetic energy of the returning debris is dissipated efficiently, (2) the material rapidly circularizes, and (3) the viscous timescale in the formed disc is short compared to the fallback time (\citealt{cannizzo90}; the results of \citet{Mockler_2018} also suggest that viscous delays are very small -- at least for UV/optical TDEs -- over timescales of hundreds of days). One can estimate the fallback rate from a full disruption of a star of mass $M_{\star}$ and radius $R_{\star}$ by an SMBH of mass $M_{\bullet}$ by assuming that the star is ``disrupted'' once it crosses the tidal radius $r_{\rm t} = R_{\star}(M_{\bullet}/M_{\star})^{1/3}$ of the black hole, whereafter the gas parcels comprising the debris stream orbit purely in the (static) Keplerian potential of the SMBH. The early behavior of the fallback rate, and in particular the rise and the peak, can be influenced by stellar properties \citep{LKP_2009, Guillochon_2013}; however, the fallback rate for these full disruptions will always asymptote to $t^{-5/3}$ at late enough times provided that there is some mass at the marginally-bound radius \citep{CN_2019}.

This model, in which the energies of the fluid elements of the star are ``frozen-in" at the tidal radius, is a reasonable approximation for full TDEs \citep{LKP_2009}. However, the surviving stellar core present in a partial TDE interacts gravitationally with the returning debris, and thereby introduces a time dependence to the gravitational potential experienced by the debris. The very existence of a conserved, Lagrangian energy for each individual gas parcel within the stream is thus violated in partial TDEs, and this approach cannot be used to self-consistently recover the late-time scaling of the fallback rate in such encounters. In addition, the core of the star in a partial TDE never actually reaches the tidal radius and it is therefore unclear how to define the radius at which the energy of the star is frozen-in.

Recently, \citet{CN_2019} developed a distinct model for analytically calculating the asymptotic temporal scaling of the fallback rate at late times from a partial TDE. By using only the conservation of mass and the Lagrangian equation of motion of the fluid elements within the stream, they concluded that the fallback rate from a partial TDE asymptotically scales approximately as $t^{-9/4}$ -- effectively independent of the mass of the core that survives the encounter (see Figure 1 of \citet{CN_2019}, which shows the asymptotic temporal power-law index of the fallback rate as a function of the mass of the surviving core). This model, however, makes a number of approximations, and ignores the pressure and self-gravity of the debris stream and the fact that the surviving core may not exactly follow a parabolic trajectory.

Here we test the validity of these approximations with hydrodynamical simulations of partial TDEs; overall we find excellent agreement between the predictions of \citet{CN_2019} and our simulations. We also define a ``break timescale," at which the fallback rate transitions from a $t^{-5/3}$ scaling to a $t^{-9/4}$ scaling, and describe how this break timescale depends on the impact parameter $\beta = r_{\rm t}/r_{\rm p}$ (where $r_{\rm p}$ is the pericenter distance of the star to the black hole), and the mass of the surviving core. We describe our simulations in Section \ref{sec:sims}, present our results in Section \ref{sec:results}, and summarize and conclude in Section \ref{sec:sum}.

\section{Simulations}\label{sec:sims}

Using the smoothed particle hydrodynamics code {\sc phantom} \citep{Price_2018}, we ran 13 simulations of the disruption of a $1 M_{\odot}$, $\gamma = 5/3$ polytropic star by a $10^6 M_{\odot}$ SMBH. Each star is set on a parabolic orbit with an impact parameter $\beta$ between 0.55 and 0.9; these constraints were found by \citet{Guillochon_2013} and \citet{Mainetti_2017} to be the approximate lower and upper bounds on $\beta$, respectively, for which partial disruptions occur for $\gamma = 5/3$ polytropes (i.e., $\beta \gtrsim 0.9$ results in a full disruption, while $\beta \lesssim 0.55$ results in no mass loss). Simulations with $\beta$ = 0.55 through 0.85 were run to $\sim$ 10 years post-disruption; simulations with $\beta =$ 0.65, 0.7, 0.75 and 0.8 were run with $10^6$ particles, while those with 0.55, 0.6 and 0.85 were run with $10^7$ particles. For reasons described below, the $\beta = 0.9$ simulation ran to 5000 years post-disruption and with $10^7$ particles.

As described in more detail in Section 4.1 of \citet{Coughlin_2016_relax}, the polytropic star is initially ``relaxed'' in isolation (i.e., without the gravitational influence of the SMBH) for ten sound crossing times over the radius of the star, which smooths out density fluctuations that arise from the numerical method. The center of mass of the relaxed polytrope is then placed at $5 r_{\rm t}$ with a velocity appropriate to the Keplerian orbit that has a pericenter distance of $r_{\rm p}$, and each particle within the star is given that same linear velocity. After the star and all of the disrupted material have passed through pericenter and subsequently reached a radius $> 5r_{\rm t}$, we excise the inner $5 r_{\rm t}$ of the computational domain, and any particles that fall through this ``accretion'' radius are ``accreted'' and contribute to the fallback rate (i.e., the rate at which particles enter this radius as a function of time); we therefore do not attempt to capture the circularization and disc formation that physically occurs. The additional numerical methodologies and physical parameters included in each simulation (e.g., the implementation of self-gravity) are identical to those in \citet{coughlin15}.

The presence of the surviving core in partial TDEs significantly increases the computational strain of the simulations: the hydrostatic nature of the surviving core limits the time step to small values relative to the dynamical time of the most bound debris. Since we are interested in the fallback rate onto the black hole, rather than the hydrodynamics of the core itself, we replace the surviving core with a point mass to make our simulations more numerically tractable. We make this replacement after a distinct core has formed, being roughly 1 day post-disruption for all but the $\beta$ = 0.9 simulation. Since the $\beta = 0.9$ simulation straddles the line between full and partial TDE, we replaced the core with a sink particle after $\sim50$ days post-disruption. Additional details of how this replacement is done numerically can be found in \citet{Golightly_2019}; we have also verified that the time at which we introduce the sink particle does not in any way affect the fallback curve, which is consistent with the fact that higher order moments of the potential of the core are negligible for parts of the stream that are not bound to the core.

\begin{figure}
    \centering
    \includegraphics[width=\linewidth]{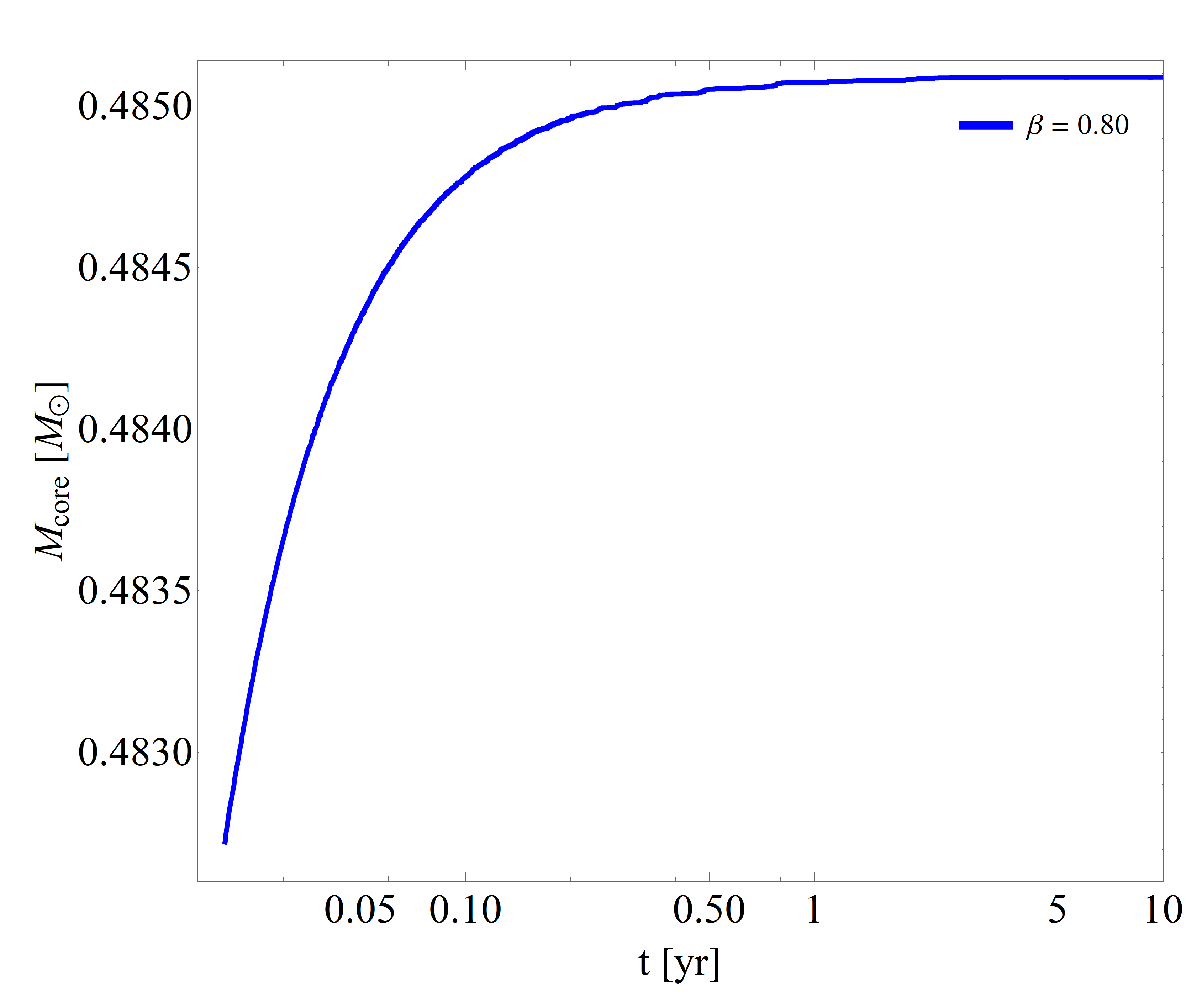}
    \caption{The mass of the surviving core as a function of time for the $\beta = 0.80$ simulation. The core mass is essentially constant for the entire simulation, increasing by less than .25\% over 10 years.}
    \label{fig:coremass}
\end{figure}

After the sink particle is created, its mass asymptotes quickly to a constant value. This behavior is illustrated in Figure \ref{fig:coremass}, in which we plot the core mass $M_{\rm core}$ as a function of time for the $\beta = 0.80$ simulation: the core mass changes by only 0.0025 $M_{\odot}$ in 10 years after creation. The core masses listed in Table \ref{tab:fulldata} (note that $\mu$ is the ratio of the mass of the surviving core to the initial mass of the star) are taken at the end of each simulation. In the Appendix, we present the results of $\beta = 0.55$ and $\beta = 0.90$ simulations that are identical aside from particle number, and show that numerical resolution has effectively no impact on our results.

\section{Results}\label{sec:results}

\subsection{Asymptotic Temporal Scaling of the Fallback Rate}

\begin{figure*}[h]
    \begin{subfloat}
        \centering
        \includegraphics[width=\linewidth]{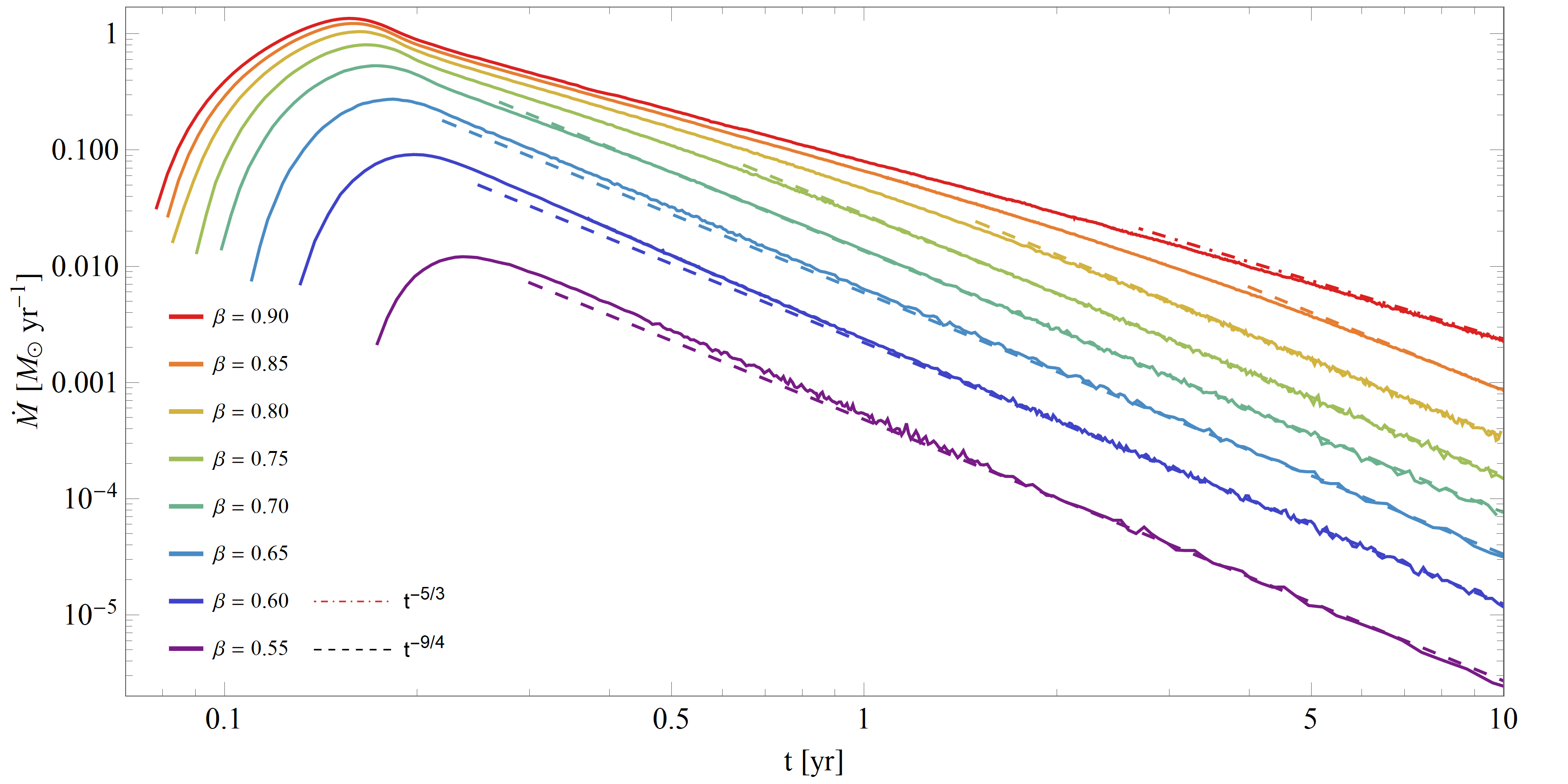}
        \label{fig:fbr}%
    \end{subfloat}%
    \begin{subfloat}
        \centering
        \includegraphics[width=\linewidth]{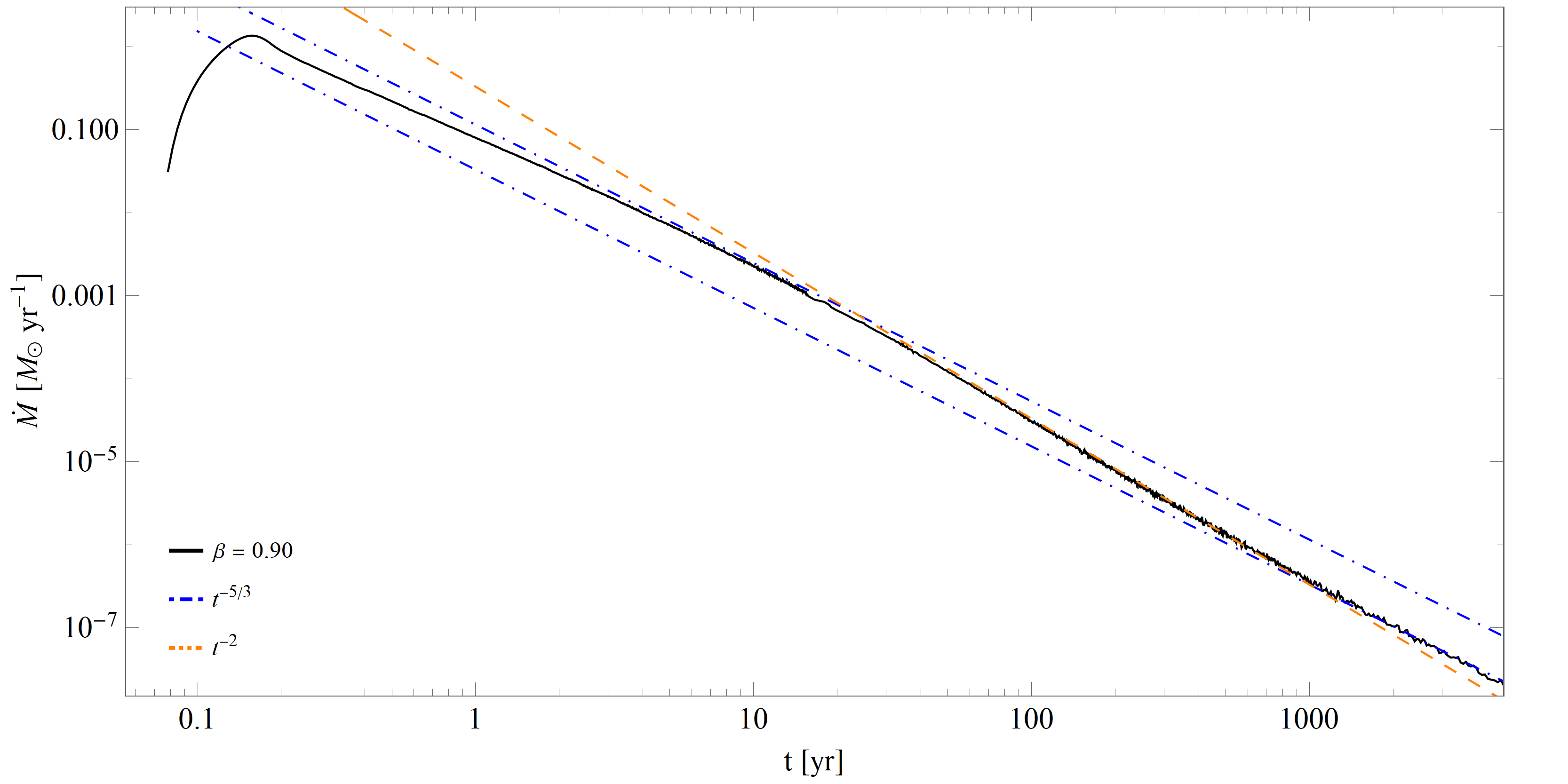}
        \label{fig:b9}%
    \end{subfloat}%
    \caption{Top: Fallback rate vs. time for eight tidal disruptions of a $1 M_{\odot}$, $\gamma = 5/3$ polytrope by a $10^6 M_{\odot}$ SMBH. These simulations span a range of $\beta$ from 0.55 (bottom, purple) to 0.9 (top, red)  as shown, and were run directly out to 15 years. $\beta = 0.55$ through $0.85$ clearly asymptote to $t^{-9/4}$, while $\beta = 0.9$ appears to asymptote to $t^{-5/3}$ by $t \sim 10$ years. For $\beta$ between $0.55-0.85$ we indicate the $t^{-9/4}$ scaling via the colored dashed lines; a $t^{-5/3}$ scaling is indicated for $\beta = 0.9$ by the red dot-dashed line. Bottom: Fallback rate for the $\beta = 0.9$ TDE, run to 5000 years post-disruption. We see that at $t \sim 30$ years the fallback rate begins to steepen, eventually approaching a temporal scaling of about $t^{-2}$ (dashed orange line), but later returns to $t^{-5/3}$ scaling (dot-dashed blue line) at very late times.}
    \label{fig:data}
\end{figure*}

The top panel of Figure \ref{fig:data} shows the fallback rate as a function of time for the $\beta$ given in the legend (solid lines represent the rates from the simulations, dashed lines give the $t^{-9/4}$ scaling, and the dot-dashed line is the canonical $t^{-5/3}$ scaling). The fallback rates for simulations with $\beta \leq 0.65$ fall off slightly steeper than $t^{-9/4}$ just after peak, but then asymptote back to $t^{-9/4}$ at later times. For deeper encounters in which $\beta$ is between 0.70 and 0.85, the fallback rates post-peak first exhibit an approximately $t^{-5/3}$ scaling before eventually steepening to a $t^{-9/4}$ scaling where they remain. We therefore observe clear adherence of all but the $\beta = 0.9$ simulation to the $t^{-9/4}$ asymptotic fallback rate temporal scaling predicted by \citet{CN_2019}. Moreover, this plot demonstrates that the asymptotic temporal scaling of the fallback rate is effectively independent of the core mass; different $\beta$ produce surviving cores of different masses (see Table \ref{tab:fulldata}), yet all simulations of $\beta \leq 0.85$ asymptote to $t^{-9/4}$ scaling, as also suggested by the analysis in \citet{CN_2019}.

On the other hand, the $\beta = 0.9$ does not seem to approach a $t^{-9/4}$ scaling by $\sim 10$ years, despite having a distinct core of about $13\%$ the mass of the original star, and instead appears to asymptote to the canonical $t^{-5/3}$ decline. Investigating the behavior of this simulation on longer timescales, however, reveals rather interesting behavior: the bottom panel of Figure \ref{fig:data} shows the fallback rate of the $\beta = 0.9$ run out to $\sim 5000$ years post-disruption. We see that between 10 and 30 years the fallback rate roughly tracks a $t^{-5/3}$ decline. It then begins to steepen, ostensibly confirming the predictions of \citet{CN_2019}, but only reaches approximately a $t^{-2}$ decline which it maintains for roughly 200 years. The fallback rate then flattens \emph{again} and eventually returns to a $t^{-5/3}$ decay by $t \sim 1000$ years.

The origin of this somewhat perplexing behavior can be understood as follows: in the model of \citet{CN_2019} the surviving stellar core follows exactly a parabolic orbit. With this assumption, both the position of the core and the marginally bound radius within the stream scale identically with time asymptotically (i.e., as $\propto t^{2/3}$), and therefore the late-time temporal power-law index of the fallback rate is independent of time and equal to $\simeq -9/4$ (see the analysis in Section 2.1 of \citealt{CN_2019}). In general, however, the energy of the center of mass is slightly modified by nonlinearities that are not contained in the tidal approximation; for mild disruptions (i.e., small $\beta$) these modifications are extremely small and leave the binding energy of the core effectively unaltered, but they become more substantial as the encounter becomes increasingly disruptive, with the surviving core being placed on a slightly hyperbolic trajectory \citep{manukian13,gafton15}.

\begin{figure}
    \centering
    \includegraphics[width=\linewidth]{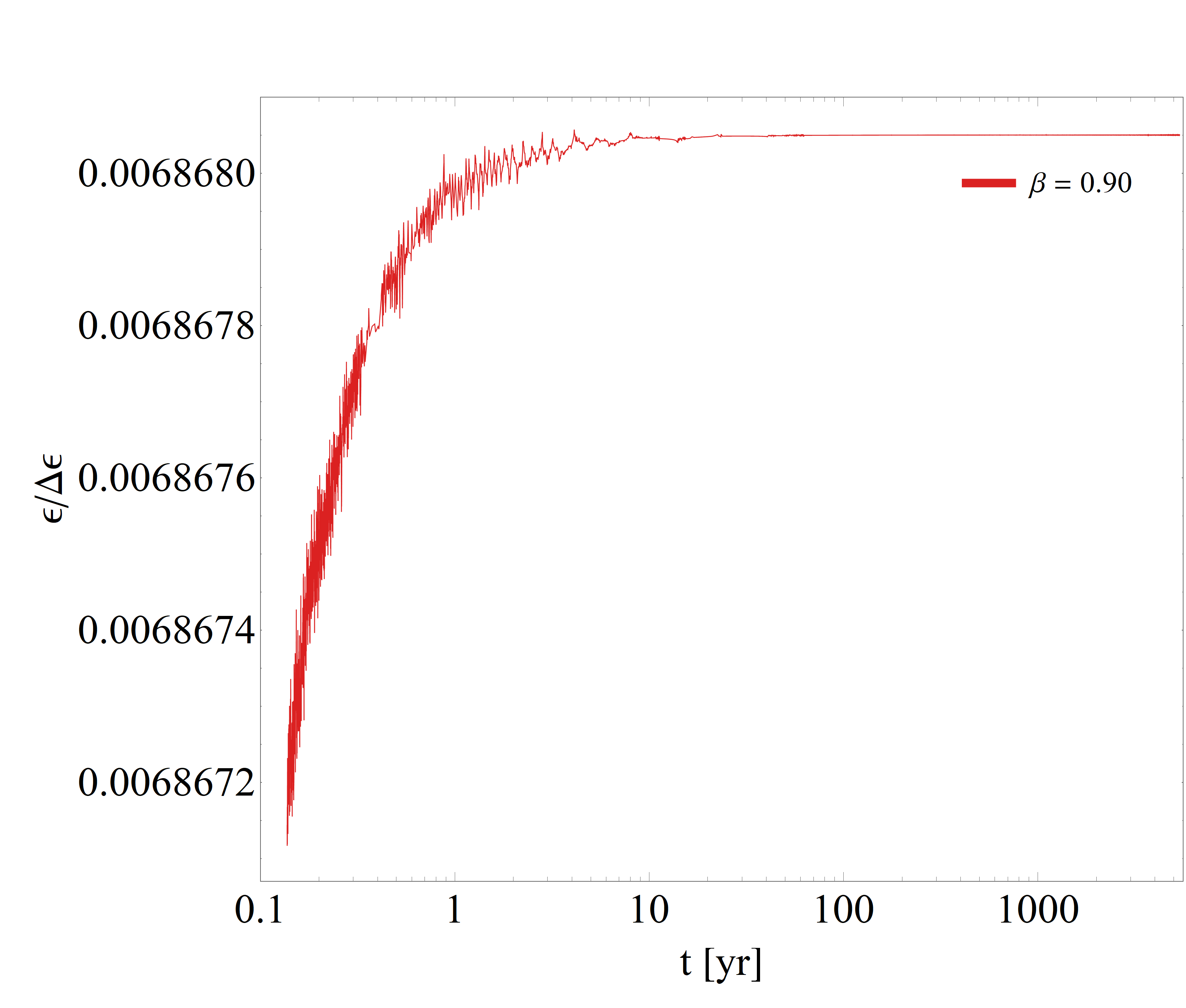}
    \caption{The specific energy of the core in the $\beta = 0.9$ simulation, normalized by the canonical energy spread $\Delta \epsilon = GM_{\bullet}R_{\star}/r_{\rm t}^2$. The energy of the core is slightly positive and effectively constant for the duration of the simulation, and for this reason the extremely late-time fallback rate for this simulation (see the bottom panel of Figure \ref{fig:data}) returns to a $t^{-5/3}$ decay.}
    \label{fig:b9coreE}
\end{figure}

For the $\beta = 0.9$ encounter, the core is initially completely disrupted by the black hole and re-forms post-disruption, which indicates that the binding energy of the core may be more substantially modified from its parabolic value in this case. Figure \ref{fig:b9coreE} illustrates the Keplerian energy of the sink particle (i.e., the orbital energy of the sink in the potential of the SMBH) as a function of time normalized by the canonical spread in the energy imparted by the tidal force, $\Delta \epsilon = GM_{\bullet}R_{\star}/R_{\rm t}^2$; $\Delta\epsilon$ is the energy that the most unbound debris would have under the impulse approximation if the star were completely disrupted at the tidal radius \citep{lacy82, LKP_2009, stone13}, and thus gives a relevant energy scale by which to normalize the binding energy of the core. This figure demonstrates that while the energy of the core is small relative to the most unbound debris and approximately equal to zero, the core is nonetheless placed on a hyperbolic orbit.

Because of the slightly hyperbolic nature of the orbit of the core, its influence on the dynamics of the marginally bound material within the stream -- which ultimately yields the deviation from the $t^{-5/3}$ power-law -- decreases over time. In particular, the \emph{initial} spatial proximity of the marginally bound radius and the core (and the fact that they both approximately follow $\propto t^{2/3}$) induces the break in the fallback rate exhibited around $\sim 30$ years in the bottom panel of Figure \ref{fig:data}, and the gravitational field of the core does have an effect on the stream dynamics. However, the unbound nature of the core implies that the distance between the marginally bound radius within the stream and the core gradually increases relative to the position of the marginally bound radius itself. For this reason, the power-law decline following the break never quite steepens to the predicted value of $t^{-9/4}$, and instead reaches a maximum decline rate that is closer to $\propto t^{-2}$. The additional distancing between the core and the marginally bound radius further flattens the fallback rate, and at sufficiently late times the core no longer influences the marginally bound material within the stream, resulting in the return to a $t^{-5/3}$ scaling by $\sim 10^3$ years\footnote{It is clear that these times are all so late that we do not expect the accretion rate onto the black hole to track the fallback rate (e.g., \citealt{cannizzo90}), and hence these findings are not immediately observationally relevant; this discussion merely provides physical understanding as to the enigmatic behavior of this fallback rate.}.

\begin{figure}[h]
    \centering
    \includegraphics[width=\linewidth]{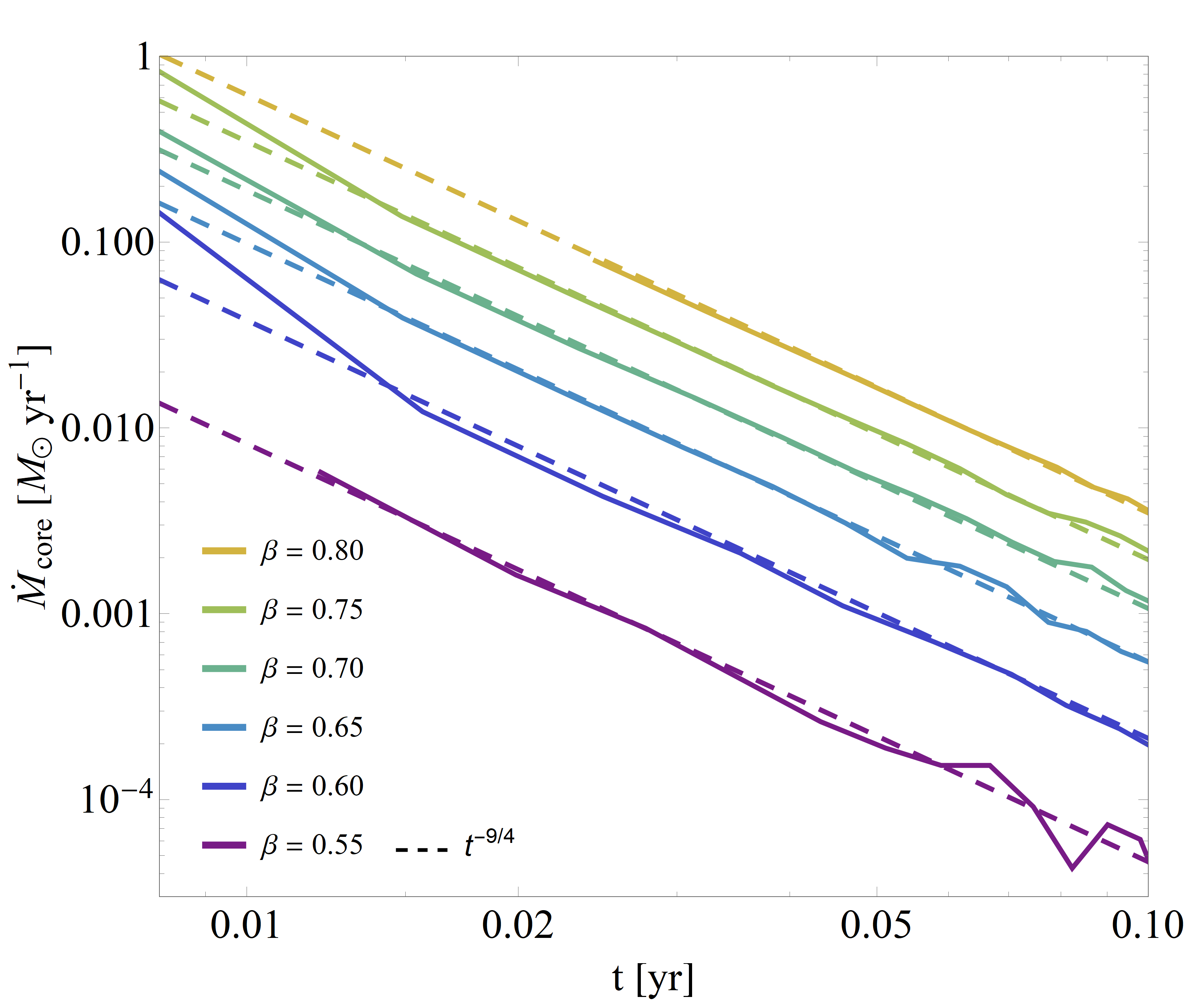}
    \caption{The fallback rate of debris to the surviving stellar core for simulations with $\beta = 0.55 - 0.80$, showing clear $t^{-9/4}$ temporal scaling.
    }
    \label{fig:cmdot}
\end{figure}

While it was not mentioned explicitly in \citet{CN_2019}, the accretion rate onto the core itself should also scale as $\propto t^{-9/4}$ within their model, which follows mathematically and straightforwardly from the analysis in Section 2.1 of \citet{CN_2019}\footnote{This can also be understood physically by noting that at late times the rate at which a fluid element leaves the marginally bound radius (in either the bound or unbound segment of the stream) cannot depend on whether it is initially slightly bound or unbound to the core. In other words, the late-time fallback rate can be thought of as accretion from a Hill sphere onto a point mass, and it does not matter which point mass we consider.}. Figure \ref{fig:cmdot} shows the fallback rates of debris to the core in simulations with $\beta = 0.55 - 0.80$, all of which are well-fit by a $t^{-9/4}$ decay. At times later than $t = 0.1$ years there is very little mass remaining within the Hill sphere of the core owing to its proximity to the core itself, which is responsible for the somewhat noisy behavior of the curves at later times in this figure.

\newpage
\subsection{Break Timescale}

\begin{figure}[h]
    \centering
    \includegraphics[width=0.495\textwidth]{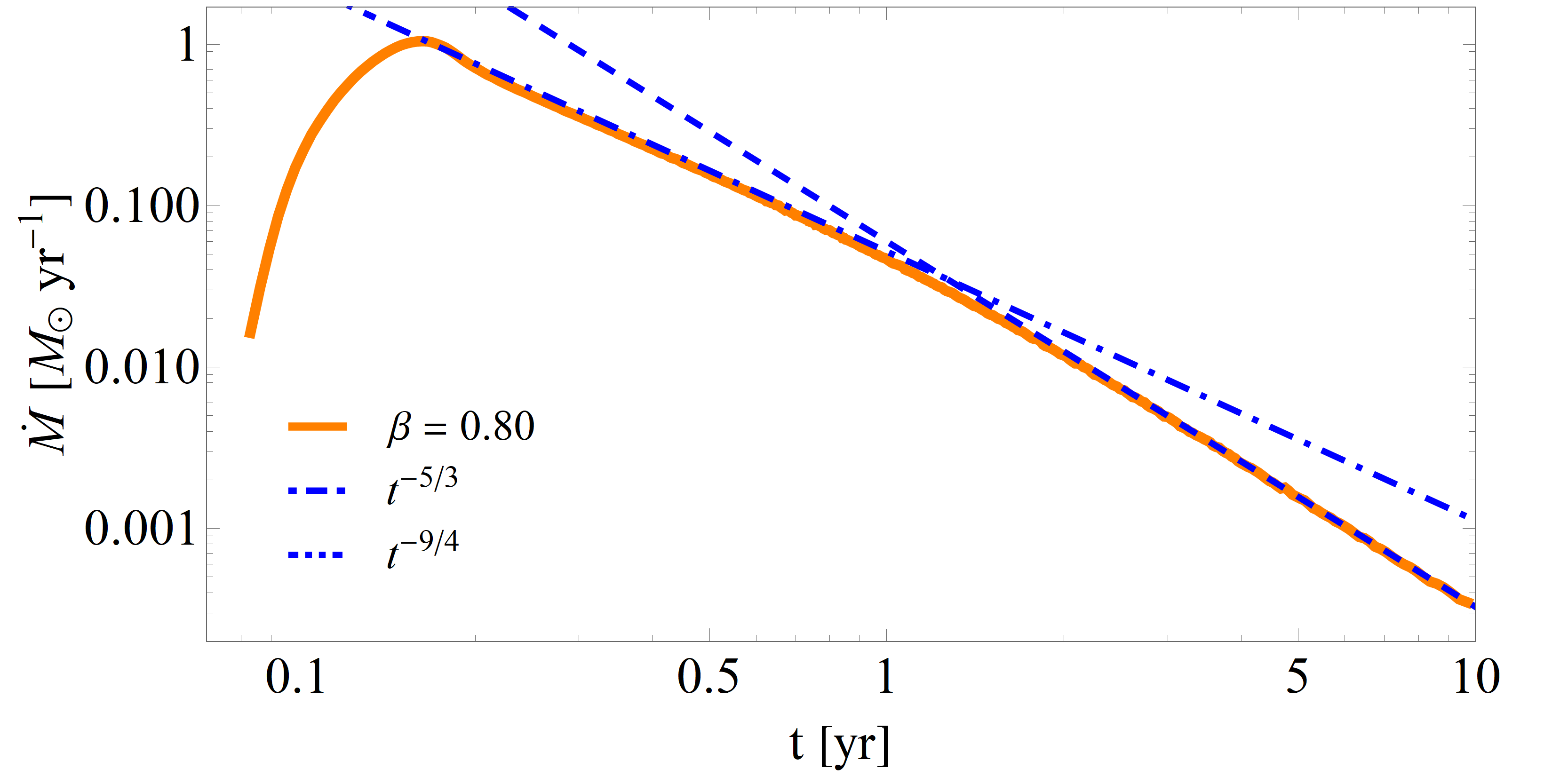}
    \caption{An example of the procedure used to calculate the break timescale: the orange solid curve illustrates the fallback rate from the $\beta = 0.8$ simulation, the dot-dashed line is a $t^{-5/3}$ curve scaled in magnitude by the peak in the fallback rate and extended forward in time, and the dotted line is a $t^{-9/4}$ curve scaled to the late-time fallback rate and extended backward in time from the latest data point. The intersection between the $t^{-5/3}$ and $t^{-9/4}$ lines delimits the break timescale. }
    \label{bt_example}
\end{figure}

We see in the top panel of Figure \ref{fig:data} that the fallback rate for partial TDEs with $\beta \leq 0.85$ always asymptotes to $t^{-9/4}$ temporal scaling effectively independently of the core mass. The core mass does, however, influence the time at which the fallback rate transitions from a decline more closely matched by a $t^{-5/3}$ scaling to a $t^{-9/4}$ scaling; clearly partial TDEs with more massive cores, corresponding to lower $\beta$, transition more rapidly to a $t^{-9/4}$ asymptotic scaling. We define this transition time -- the ``break timescale" $t_{\rm break}$ -- as follows: for a given fallback curve, extend a $t^{-5/3}$ line forward in time from the peak, and extend a $t^{-9/4}$ line backwards from the late-time, $t^{-9/4}$ portion of the curve. The break timescale $t_{\rm break}$ is then the time at which these two lines intersect. An example of this procedure for the $\beta = 0.8$ data is shown in Figure \ref{bt_example}. 

\begin{figure}[h]
    \centering
    \includegraphics[width=\linewidth]{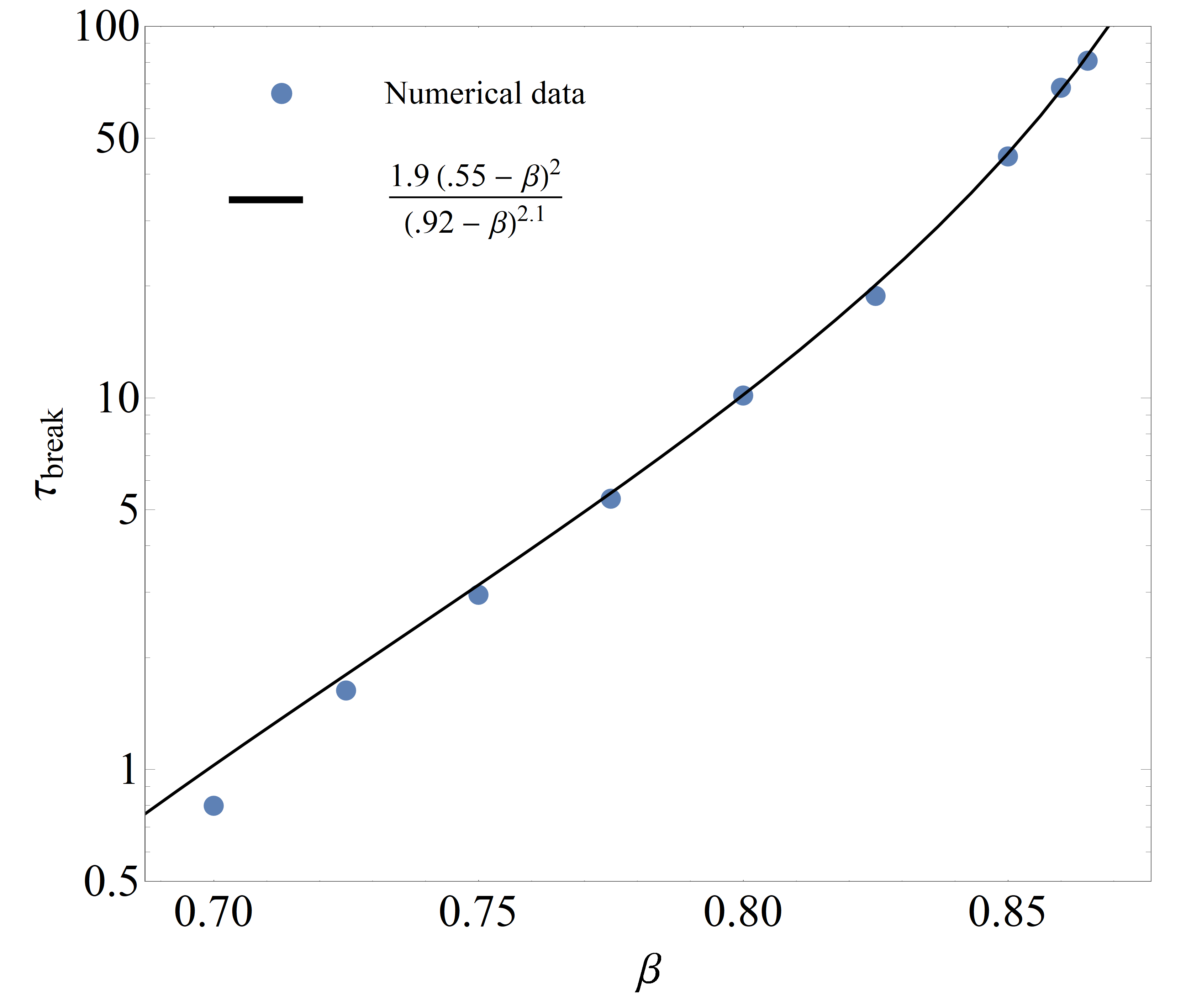}
    \caption{A log-linear plot of $\tau_{\rm break}$ as a function of the impact parameter $\beta$. We see that the fallback rates of grazing encounters of the star with the SMBH are fastest to approach their asymptotic temporal scaling.}
    \label{fig:btvsbeta}
\end{figure}

\begin{figure}[h]
    \centering
    \includegraphics[width=\linewidth]{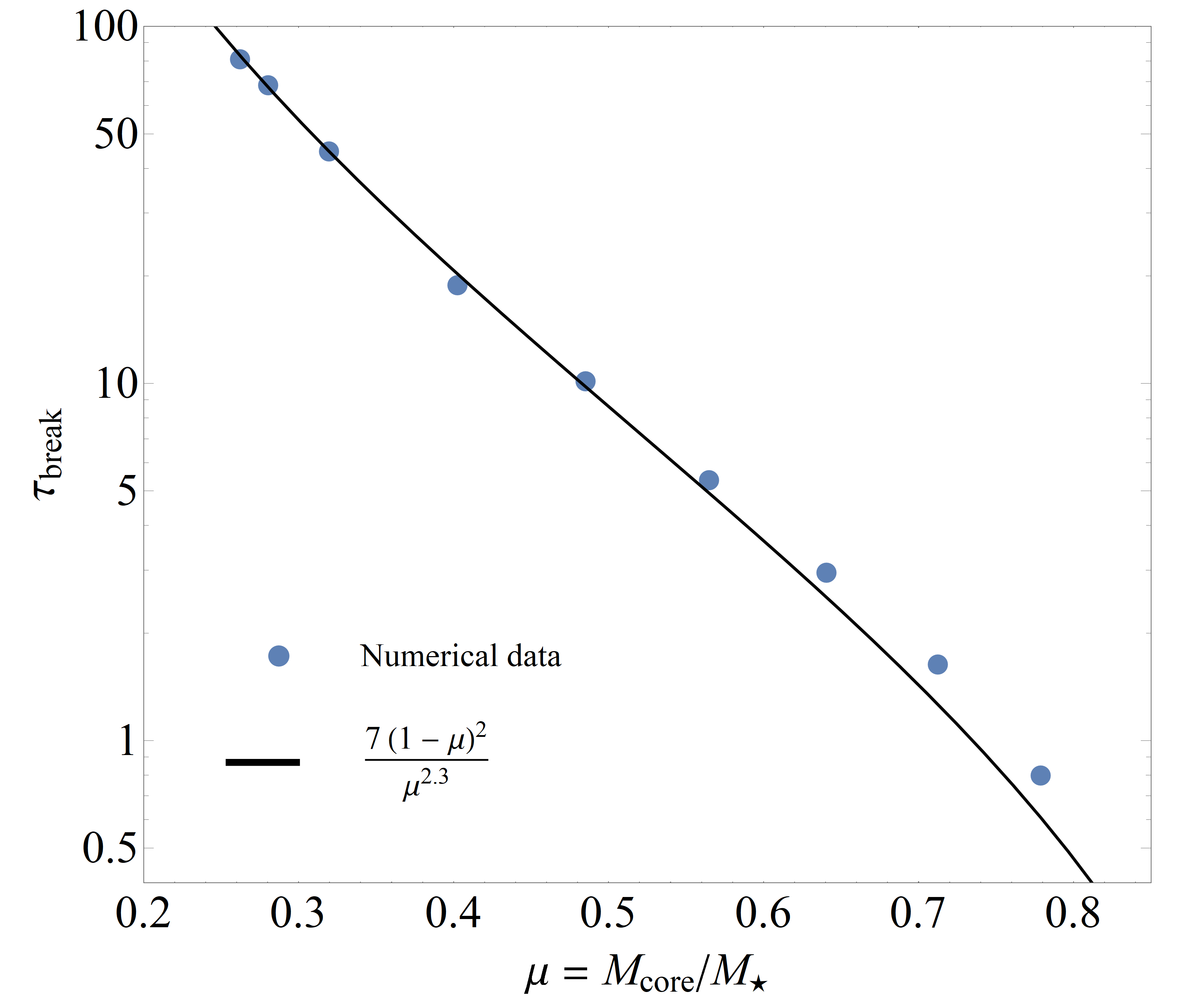}
    \caption{A log-linear plot of $\tau_{\rm break}$ as a function of the fractional core mass $\mu = M_{\rm core}/M_{\star}$. Partial TDEs with more massive cores approach their asymptotic fallback rate temporal scaling more quickly than those with less massive cores.}
    \label{fig:btvscm}
    \vspace{-5mm}
\end{figure}

In Figures \ref{fig:btvsbeta} and \ref{fig:btvscm} and Table \ref{tab:fulldata}, we study the dimensionless break timescale $\tau_{\rm break}$, which we calculate by subtracting the time to peak of the fallback rate from $t_{\rm break}$ and dividing by the return time of the most-bound debris under the impulse approximation, $T_{\rm mb}$, given by (e.g., \citealt{LKP_2009})
\begin{equation}
    T_{\rm mb} = \left( \frac{R_{\star}}{2}\right)^{3/2} \frac{2\pi M_{\bullet}}{M_{\star}\sqrt{GM_{\bullet}}}.
    \label{eqn:Tmb}
\end{equation}
$T_{\rm mb}$, which for our simulation parameters is $\sim$ 41 days, encapsulates the overall dependence of any specific timescale associated with a TDE (e.g., the time to peak) on the bulk stellar properties (for a given stellar structure, e.g., a $\gamma = 5/3$ polytrope) and black hole mass. Thus, as argued by \citet{LKP_2009}, \citet{Guillochon_2013}, \citet{Mockler_2018}, and \citet{Golightly_2019}, dividing the break timescale measured from the simulation by $T_{\rm mb}$ should remove the bulk dependence on the black hole mass and the stellar properties (again, for a given stellar structure; this was also shown directly by \citealt{Wu_2018} for the black hole mass dependence). We therefore expect that the ratio $t_{\rm break}/T_{\rm mb}$ is only a function of the impact parameter $\beta$ and the mass of the stellar core $M_{\rm core}$. If a physical break timescale $t_{\rm break}$ is observed, we can infer the $\beta$ of the encounter if we assume a set of stellar properties and the black hole mass is known by other means (e.g., through a black hole mass scaling relationship).

Figure \ref{fig:btvsbeta} shows $\tau_{\rm break}$ as a function of $\beta$.
To fit our data we impose two physical constraints: from the lower (upper) bound on $\beta$ for which a core will survive a TDE \citep{Mainetti_2017}, we require $\tau_{\rm break}$ to go to zero (infinity) at $\beta =$ 0.55 (0.92). We find that the relationship $\tau_{\rm break} = 1.9\left(.55 - \beta\right)^2/\left(0.92 - \beta\right)^{2.1}$ provides a simple and reasonable interpolation of the data, as shown by the black curve in this figure. In general, partial TDEs in which the star dives deeper into the tidal field of the SMBH yield fallback rates that more slowly reach their asymptotic, $t^{-9/4}$ scaling than those from more grazing encounters. 

Figure \ref{fig:btvscm} shows $\tau_{\rm break}$ as a function of $\mu$, where $\mu = M_{\rm core}/M_{\star}$ is the ratio of the surviving core mass (at the end of the simulation) to the initial stellar mass\footnote{Note that, in the analytic model of \citet{CN_2019}, the quantity that manifestly determines the asymptotic fallback rate is the ratio of the mass of the core to the mass of the supermassive black hole. For this reason, \citet{CN_2019} introduced $\mu_{-6} = M_{\rm core}/M_{\bullet}\times10^{6}$, which is identical to the ratio of the core mass to the initial stellar mass (for this setup) but maintains the physicality of the black hole dependence; to avoid unnecessarily cumbersome notation here, we simply define $\mu = M_{\rm core}/M_{\star}$ without reference to the black hole mass.}, alongside the fit $\tau_{\rm break} = 7\left(1-\mu\right)^{2}/\mu^{2.3}$. As for the constraints on the relationship between $\tau_{\rm break}$ and $\beta$, this simple functional form is consistent with the notions that the break timescale should go to infinity when there is no surviving core ($\mu = 0$) and should be small when there is little mass lost from the core ($\mu \simeq 1$). In general, the fallback rate for partial TDEs with more massive cores reaches an asymptotic $t^{-9/4}$ scaling more quickly than those with less massive cores, which is consistent with the predictions of \citet{CN_2019} (see their Figure 2).

\begin{table}[h]
    \centering
    \vspace{4mm}
        \begin{tabular}{c|c|c|c}
            $\beta = r_{\rm t}/r_{\rm p}$ & $\mu = M_{\rm core}/M_{\star}$ & $\tau_{\rm break}$ & $t_{\rm break}$ [yr] \\
            \hline
            0.550 & 0.994 & -- & -- \\
             0.600 & 0.962 & -- & -- \\
             0.650 & 0.889 & -- & -- \\
            0.700 & 0.779 & 0.799 & 0.263 \\
            0.725 & 0.712 & 1.633 & 0.352 \\
            0.750 & 0.641 & 2.954 & 0.497 \\
            0.775 & 0.565 & 5.357 & 0.763 \\
            0.800 & 0.485 & 10.14 & 1.300 \\
            0.825 & 0.402 & 18.83 & 2.269 \\
            0.850 & 0.320 & 44.63 & 5.166 \\
            0.860 & 0.280 & 68.37 & 7.822 \\
            0.865 & 0.262 & 80.88 & 9.225 \\
            0.900 & 0.135 & -- & -- \\
            \hline
        \end{tabular}
    \caption{A table of $\beta = r_{\rm t}/r_{\rm p}$ with $r_{\rm t}$ the tidal radius and $r_{\rm p}$ the pericenter distance, $\mu = M_{\rm core}/M_{\star}$ with $M_{\rm core}$ the mass of the surviving core and $M_{\star}$ the initial stellar mass, $\tau_{\rm break}$ the dimensionless break time at which the fallback rate transitions from a $\sim t^{-5/3}$ decline to a $t^{-9/4}$ decline, and $t_{\rm break}$ the break time measured from the simulation in units of years.}
    \label{tab:fulldata}
    \vspace{-2mm}
\end{table}

We include our data for $\beta$, $\mu$, $\tau_{\rm break}$ and $t_{\rm break}$ in Table \ref{tab:fulldata} below. The simulations for which $\beta < 0.7$ do not follow a $t^{-5/3}$ scaling for any significant length of time, and therefore do not exhibit an associated break timescale. The simulations for which $\beta > 0.865$, except $\beta = 0.9$, ran for only 10 years post-disruption; this was not long enough for the fallback rate to reach a stable asymptotic temporal scaling, so we could not measure $\tau_{\rm break}$ for these simulations. The $\beta = 0.9$ simulation ran to over 5000 years -- plenty of time for the fallback rate to exhibit a clear break in the temporal scaling from $t^{-5/3}$ -- but the fallback rate only steepens to approximately a $t^{-2}$ temporal scaling, not $t^{-9/4}$ as our definition of the break time requires (see the text for further discussion). We therefore do not give a break time for the $\beta = 0.9$ simulation.

\newpage
\section{Summary and Conclusions}\label{sec:sum}

In this paper we presented the results of a set of SPH simulations of partial tidal disruptions of a $1\, M_{\odot}$, $\gamma = 5/3$ polytrope by a $10^6 M_{\odot}$ SMBH. We observed that partial TDEs with $\beta$ between 0.55 and 0.85 -- nearly the entire range of $\beta$ for which partial TDEs can occur for $\gamma = 5/3$ polytropes \citep{Mainetti_2017, Guillochon_2013} -- asymptote to a $t^{-9/4}$ temporal scaling. Since each $\beta$ corresponds to a unique core mass for a given star (see Table \ref{tab:fulldata}), these results suggest that the asymptotic temporal scaling of partial TDE fallback rates is effectively independent of the mass of the surviving core.

We also defined a ``break timescale,'' the time at which the fallback rate transitions from $t^{-5/3}$ scaling to $t^{-9/4}$ scaling. We found that the dimensionless break timescale $\tau_{\rm break}$ (defined as the ratio of physical time to the return time of the most bound debris $T_{\rm mb}$ given in equation \ref{eqn:Tmb}) relates to $\beta$ as $\tau_{\rm break} = 1.9\left(.55 - \beta\right)^2/\left(0.92 - \beta\right)^{2.1}$, and relates to the core mass fraction $\mu$ as $\tau_{\rm break} = 7\left(1-\mu\right)^{2}/\mu^{2.3}$. If such a break in the power law is observed, it can be used to infer properties of partial TDEs such as impact parameter and surviving core mass once a given stellar structure is assumed.

We emphasize that for grazing encounters in which only a small fraction of the star's mass is successfully removed, the fallback rate will never conform to a $t^{-5/3}$ temporal scaling. This result has significant implications for the interpretation of observational data from partial TDEs. For example, \citet{Gomez_2020} analyze data from AT 2018hyz, a recently observed TDE. They find an impact parameter of $\beta =$ 0.6 and a surviving core mass of roughly 90\% the mass of the undisrupted star, yet their light curve fit yields a $t^{-5/3}$ temporal scaling. If the estimated $\beta$ and core mass fraction are accurate, then AT 2018hyz is a partial disruption and, given the results presented here (and in particular Figure \ref{fig:data}), its light curve should follow a $t^{-9/4}$ temporal scaling if the accretion luminosity is tracking the fallback rate; this conclusion is clearly discrepant with the interpretation in \citet{Gomez_2020}.

Because our main aim here was to assess the sensitivity of the predicted, $t^{-9/4}$ scaling on more realistic sets of physical conditions (than those adopted in \citealt{CN_2019}; see Section \ref{sec:intro}), we only studied the disruption of stars with a single stellar profile, being a $\gamma = 5/3$ polytrope. We plan to investigate the dependence of various physical properties of the fallback curve identified here (e.g., the break timescale and the asymptotic power-law rate) on stellar structure in future investigations.\\

\software{\sc phantom \citep{Price_2018}, Mathematica \citep{Mathematica} }

\acknowledgements
The authors thank the anonymous referee for a constructive report, and Jim Pringle for comments on the manuscript. PRM acknowledges support from the Syracuse University HTC Campus Grid and NSF award ACI-1341006. ERC acknowledges support from NASA through the Hubble Fellowship, grant No.~HST-HF2-51433.001-A awarded by the Space Telescope Science Institute, which is operated by the Association of Universities for Research in Astronomy, Incorporated, under NASA contract NAS5-26555. CJN is supported by the Science and Technology Facilities Council (grant number ST/M005917/1). Some of this work was performed using the DiRAC Data Intensive service at Leicester, operated by the University of Leicester IT Services, which forms part of the STFC DiRAC HPC Facility (www.dirac.ac.uk). The equipment was funded by BEIS capital funding via STFC capital grants ST/K000373/1 and ST/R002363/1 and STFC DiRAC Operations grant ST/R001014/1. DiRAC is part of the National e-Infrastructure.

\bibliography{main}
\bibliographystyle{aasjournal}

\appendix

\begin{figure}[h]
    \centering
    \includegraphics[width=\textwidth]{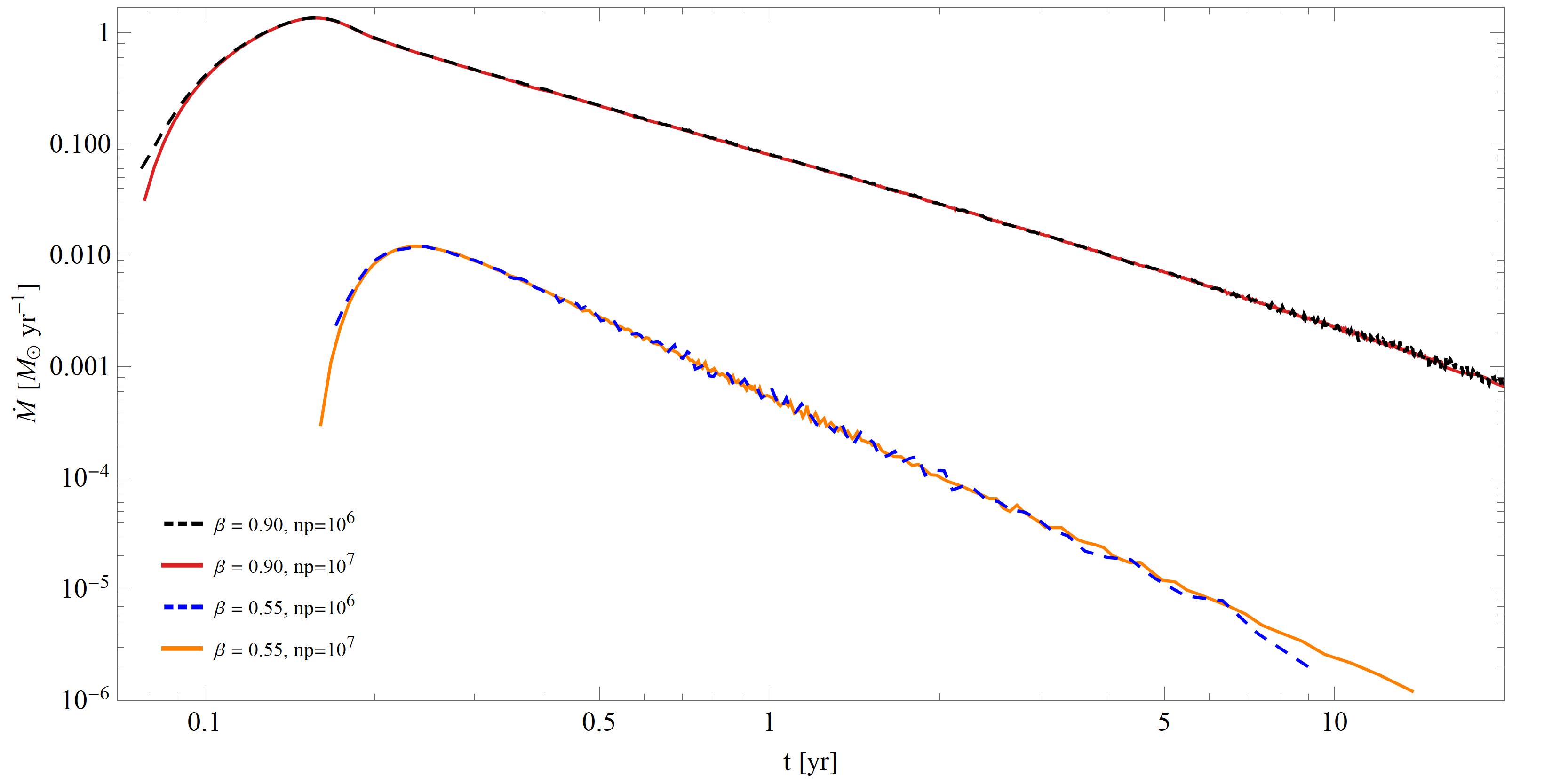}
     \caption{
     The fallback rates from simulations of $\beta=0.55$ and $\beta = 0.90$, $\gamma = 5/3$ partial TDE. The dashed black curve is $\beta=0.90$ with $10^6$ particles; the solid red curve is $\beta=0.90$ with $10^7$ particles; the dashed blue curve is $\beta=0.55$ with $10^6$ particles; and the solid orange curve is $\beta=0.55$ with $10^7$ particles. There are clearly only very small differences between the two solutions for the fallback rate, which directly demonstrates the insensitivity of our results to numerical resolution.}
    \label{appendix:num_res}
\end{figure}

As noted in Section \ref{sec:sims}, the fallback rates were calculated with either $10^6$ or $10^7$ particles. $10^7$ particles were used for simulations with $\beta$ close to the critical value for full disruption ($\beta \simeq 0.9$) or no mass loss ($\beta \simeq 0.55$), as in these cases the fallback rate either transitioned to its asymptotic decline at late times (when there were relatively few particles being accreted) or the stellar debris stream was composed of only a small percentage of the stellar mass and hence very few particles. Therefore, to obtain reasonable estimates of the fallback rate and to remove excessive levels of Poisson noise associated with the return of discrete particles, the particle number was augmented for these disruptions. 

The fallback curves in this paper were created by binning the incremental changes in the mass accreted over time as particles cross the accretion radius. At early times (typically within roughly half of the time to peak fallback rate) there is a very large flux of particles on timescales that are much shorter than the dynamical time of the most bound debris; at these early times, therefore, binning the incremental mass changes in linear time steps is sufficient to reduce the Poisson noise in the fallback rate that arises from the finite number of particles. Here the plots use a time step on the order of $\sim 0.5$ days, but changing this number by factors of a few does not change the result. At late times, however, the relatively few number of particles that have yet to be accreted implies that employing this same fixed time step induces significant noise, i.e., substantial variation in the number of particles accreted from one timestep to the next, in the calculated fallback rate. Therefore, at late times the temporal bin width over which the mass changes incrementally is calculated by requiring that a fixed number of particles be accreted, which ensures that we are consistently averaging over the same number of particles. We adopted 30 for this number in all plots presented in this paper except for the $10^6$ particle $\beta=0.55$ simulation shown above, for which we chose 10 to ensure that the curve extended past 10 years despite so few particles being accreted by this time. Changing this number by modest factors (e.g., in going from 30 to 10 or from 30 to 60) only increases the level of noise (by significantly reducing this number below 30) or reduces the level of detail (by significantly increasing this number above 30).

To assess the sensitivity of our results to the particle number employed, Figure \ref{appendix:num_res} illustrates the fallback rates for simulations identical in their physical setup -- a $\gamma = 5/3$ polytrope disrupted by a $10^6$ black hole with $\beta = 0.90$ or 0.55 -- but with either $10^6$ (dashed curves) or $10^7$ (solid curves) particles. For the $\beta=0.90$ data, the two fallback rates differ slightly in their pre-peak behavior, with the $10^6$ simulation yielding a slightly earlier return time of the most bound debris and a larger fallback rate overall compared to the $10^7$ simulation. While some noise appears in the $10^6$ simulation at late times, the post-peak behavior, including the asymptotic fallback rate and its temporal scaling, is effectively identical at every point. Due to the minuscule amount of mass liberated in the $\beta=0.55$ disruption, both the $10^6$ and $10^7$ curves exhibit more noise than the $\beta=0.90$ simulations. However, the time-to-peak and the post-peak behavior are again effectively identical at every point between the two simulations with different numerical resolution, but with a systematically larger amount of noise present in the $10^6$ particle simulation; for this simulation, since the core contains $\sim 99.4\%$ of the mass, there are only $\sim 3000$ particles contained in the returning debris stream, which is the origin of this greater degree of noise. This figure demonstrates that the results of our study are not dependent on numerical resolution.

\end{document}